\documentclass[twocolumn,showpacs,preprintnumbers,amsmath,amssymb]{revtex4}
\usepackage{graphicx}
\begin{document}

\title{Graphene-based photonic crystal}

\author{Oleg L. Berman$^{1}$, Vladimir S. Boyko$^{1}$, Roman Ya. Kezerashvili%
$^{1,2}$, Anton A. Kolesnikov$^{3}$, and Yurii E. Lozovik$^{3,4}$}
\affiliation{\mbox{$^{1}$Physics Department, New York City College
of Technology, The City University of New York,} \\
Brooklyn, NY 11201, USA \\
\mbox{$^{2}$The Graduate School and University Center, The City
University of New York,}
New York, NY 10016, USA \\
\mbox{$^{3}$ Institute of Spectroscopy, Russian Academy of
Sciences,} \\
142190 Troitsk, Moscow Region, Russia \\
 \mbox{$^{4}$Moscow
Institute of Physics and Technology (State University), 141700,
Dolgoprudny, Russia }}

\begin{abstract}

A novel type of photonic crystal formed by embedding a periodic
array of constituent stacks of alternating graphene and dielectric
discs  into a background dielectric medium is proposed.    The
photonic band structure and transmittance of such photonic crystal
are calculated. The graphene-based photonic crystals can be used
effectively as the frequency filters and waveguides for the far
infrared region of electromagnetic spectrum. Due to substantial
suppression of absorption of low-frequency radiation in doped
graphene the damping and skin effect in the photonic crystal are
also suppressed. The advantages of the graphene-based photonic
crystal are discussed.

\end{abstract}

\pacs{42.70.Qs, 78.67.Wj, 78.67.-n, 78.67.Pt}

\maketitle

Photonic crystals attract the growing interest due to various modern
applications~\cite{Eldada,Chigrin}. For example, they can be used
 as the frquency filters and waveguides. Photonic
crystals are media with a spatially periodical dielectric
function~\cite{Yablonovitch1,John,Joannopoulos1,Joannopoulos2}. This
periodicity can be achieved by embedding a periodic array of
constituent elements with dielectric constant $\varepsilon _{1}$ in
a background medium characterized by dielectric constant
$\varepsilon _{2}$.  Different materials have been used for the
corresponding constituent elements including
dielectrics~\cite{Joannopoulos1,Joannopoulos2}, semiconductors,
metals~\cite{Ulrich,Maradudin,Kuzmiak}, and
superconductors~\cite{Takeda1,Berman1,Lozovik,Berman2}.

A novel type of 2D electron system was experimentally observed in
graphene, which is a 2D honeycomb lattice of the carbon atoms that
form the basic planar structure in
graphite~\cite{Novoselov1,Lukyanchuk,Zhang1}. Due to unusual
properties of the band structure, electronic properties of graphene
became the object of many recent experimental and theoretical
studies \cite{Novoselov1,Lukyanchuk,Zhang1,
Novoselov2,Zhang2,Falko,Katsnelson,Castro_Neto_rmp}.  Graphene is a
gapless semiconductor with massless electrons and holes which have
been described as
Dirac-fermions~\cite{Novoselov1,Lukyanchuk,DasSarma}. The unique
electronic properties  of graphene
 in a magnetic field have been studied recently
 \cite{Nomura,Jain,Gusynin1,Gusynin2}.   It was shown that in infrared and at larger wavelengths
  transparency of graphene is defined by the fine structure constant~\cite{Nair}.
 Thus, graphene has unique optical properties.  The space-time dispersion of graphene
 conductivity was analyzed in Ref.~[\onlinecite{Varlamov}] and the optical properties of graphene were
  studied in  Refs.~[\onlinecite{Falkovsky_prb,Falkovsky_conf}].

As mentioned above, the photonic crystals with the dielectric,
metallic, semiconductor, and superconducting constituent elements
have different photonic band and transmittance spectra. The
dissipation of the electromagnetic wave in all these photonic
crystals is different. The photonic crystals with the metallic and
superconducting constituent elements can be used as the frequency
filters and waveguides for the far infrared region of spectrum,
while the dielectric photonic crystals can be applied for the
devices only for the optical region of spectrum.

\begin{figure}
\includegraphics[width = 3.5in]{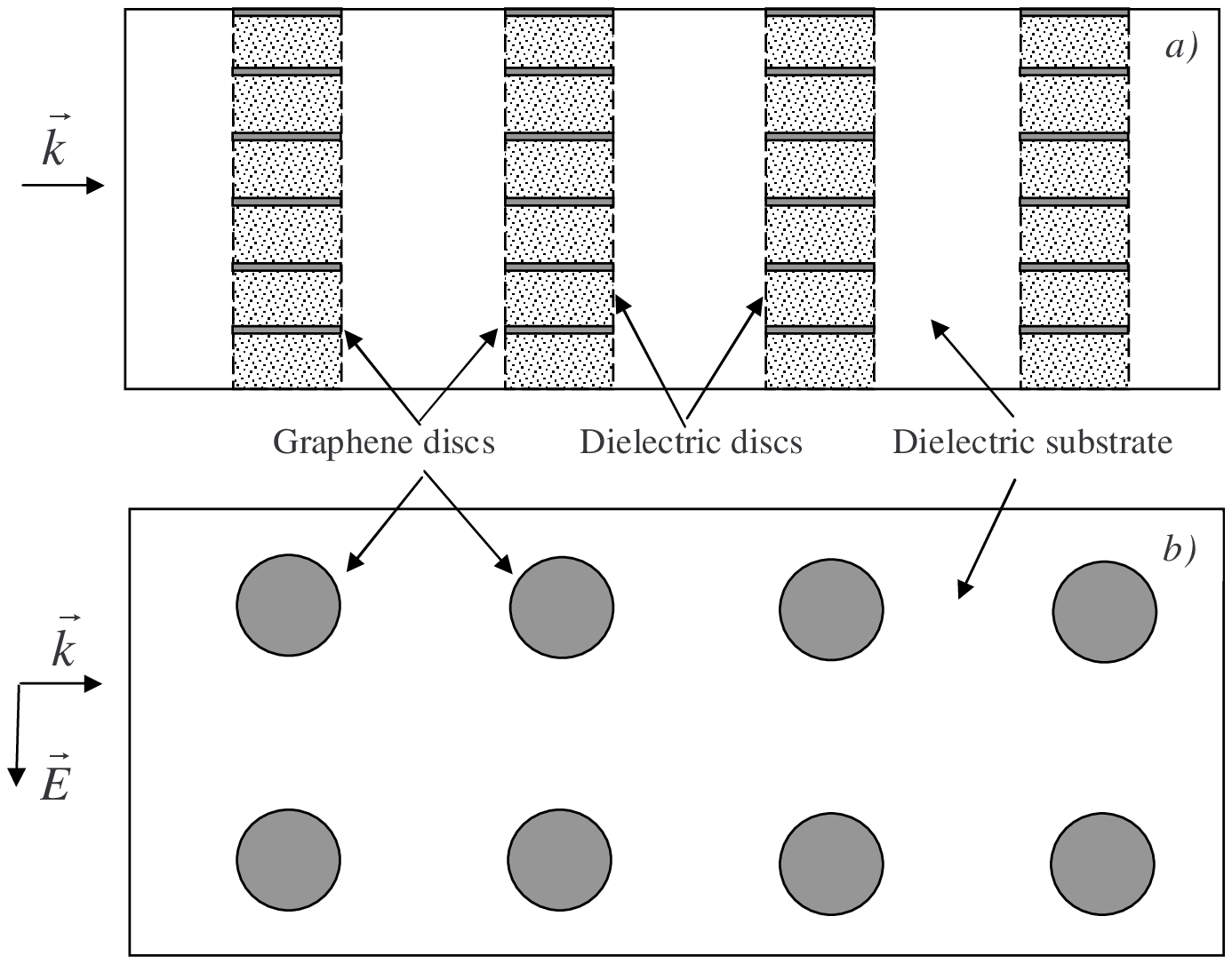}
\caption{Graphene-based photonic crystal: a) the side view.  The
material of the dielectric between graphene discs can be the same as
the material of the dielectric substrate; b) the top view.}
\label{crys}
\end{figure}

In this Letter, we consider a 2D photonic crystal formed by stacks
of periodically placed graphene discs embedded into the dielectric
film. The  stack is formed by graphene discs placed one on top of
another separated by the dielectric placed between them as shown in
Fig.~\ref{crys}. We calculate the photonic band structure and
transmittance of this graphene-based photonic crystal. We will show
that the graphene-based photonic crystals can be applied for the
devices for the far infrared region of spectrum.

We consider polarized electromagnetic waves with the electric field $\mathbf{E}$ parallel to the
plane of the graphene discs.

Expanding the electric field in terms of the Bloch waves inside a
photonic crystal, one obtains from the wave equation the system of
equations for Fourier components of the electric
field~\cite{Joannopoulos2, Maradudin}:
\begin{eqnarray}\label{fur}
(\mathbf{k} + \mathbf{G})^{2}E_{\mathbf{k}}(\mathbf{G}) =
\frac{\omega^{2}(\mathbf{k})}{c^{2}} \sum_{\mathbf{G}'}\varepsilon
(\mathbf{G}-\mathbf{G}') E_{\mathbf{k}}(\mathbf{G}')  \ ,
\end{eqnarray}
which presents the eigenvalue problem for finding photon dispersion
curves $\omega(\mathbf{k})$.  In Eq.~(\ref{fur}) the coefficients of
the Fourier expansion for the dielectric constant are given by
\begin{eqnarray}\label{diel1}
\varepsilon (\mathbf{G}-\mathbf{G}') =
\varepsilon_{0}\delta_{\mathbf{G}\mathbf{G}'} + (\varepsilon_{1} -
\varepsilon_{0})M_{\mathbf{G}\mathbf{G}'}   \ .
\end{eqnarray}
In Eq.~(\ref{diel1}) $\varepsilon_{0}$ is the dielectric constant of
the dielectric, $\varepsilon_{1}$ is the  dielectric constant of
graphene multilayers separated by the dielectric material, and
$M_{\mathbf{G}\mathbf{G}'}$ for the geometry considered above is
\begin{eqnarray}\label{mg}
M_{\mathbf{G}\mathbf{G}'} &=& 2 f \frac{J_{1}(|\mathbf{G} -
\mathbf{G}'|r)}{(|\mathbf{G} - \mathbf{G}'|r)} \ , \ \ \ \
\mathbf{G} \ne \mathbf{G}'  \ , \nonumber \\
M_{\mathbf{G}\mathbf{G}'} &=& f \ , \ \ \ \ \mathbf{G} = \mathbf{G}'
\ ,
\end{eqnarray}
where $J_{1}$ is the Bessel function of the first order, and $f =
S_{g}/S$ is the filling factor of 2D photonic crystal.

In our consideration the size of the graphene discs was assumed to
be much larger than the period of the graphene lattice, and we
applied the expressions for the dielectric constant of the infinite
graphene layer for the graphene discs, neglecting the effects
related to their finite size.

The dielectric constant $\varepsilon_{1}(\omega) $ of graphene
multilayers system separated by the dielectric layers  with the
dielectric constant $\varepsilon_{0}$ and the thickness $d$ is given
by~\cite{Falkovsky_prb,Falkovsky_conf}
\begin{eqnarray}\label{dielmult}
\varepsilon_{1}(\omega) = \varepsilon_{0} + \frac{4\pi i \sigma_{g}
(\omega)}{\omega d} \ ,
\end{eqnarray}
where $\sigma_{g}(\omega)$ is the dynamical conductivity of doped graphene for
the high frequencies ($\omega \gg kv_{F}$, $\omega \gg \tau^{-1}$)
at temperature $T$ given by~\cite{Falkovsky_prb,Falkovsky_conf}
\begin{eqnarray}\label{cond}
&& \sigma_{g}(\omega) = \frac{e^{2}}{4\hbar} \left[ \eta (\hbar
\omega - 2\mu) \right. \nonumber \\ && \left. + \frac{i}{2\pi}
\left( \frac{16 k_{B} T}{ \hbar
\omega}\log\left[2\cosh\left(\frac{\mu}{2k_{B}T}\right)\right]
\right. \right. \nonumber \\ && \left. \left. - \log
\frac{(\hbar\omega + 2\mu)^{2}}{(\hbar\omega - 2\mu)^{2} +
(2k_{B}T)^{2}}\right) \right] \ .
\end{eqnarray}
Here $e$ is the charge of an electron, $\tau^{-1}$ is the electron
collision rate, $k$ is the wavevector, $v_{F} = 10^{8} \
\mathrm{cm/s}$ is the Fermi velocity of electrons in
graphene~\cite{Falkovsky_conf}, $k_{B}$ is the Boltzmann constant,
and $\mu$ is the the chemical potential determined by the electron
concentration $n_{0} = (\mu/(\hbar v_{F}))^{2}/\pi$, which is
controlled by the doping. The chemical potential can be calculated
as $\mu = (\pi n_{0})^{1/2}\hbar v_{F}$. In the calculations below
we assume $n_{0} = 10^{11} \ \mathrm{cm^{-2}}$.  For simplicity, we
assume that the dielectric material is the same for the dielectric
discs between the graphene disks and between the stacks. As the
dielectric material we consider  SiO$_{2}$ with the dielectric
constant $\varepsilon_{0} = 4.5$.

\begin{figure}
\includegraphics[width = 3.5in]{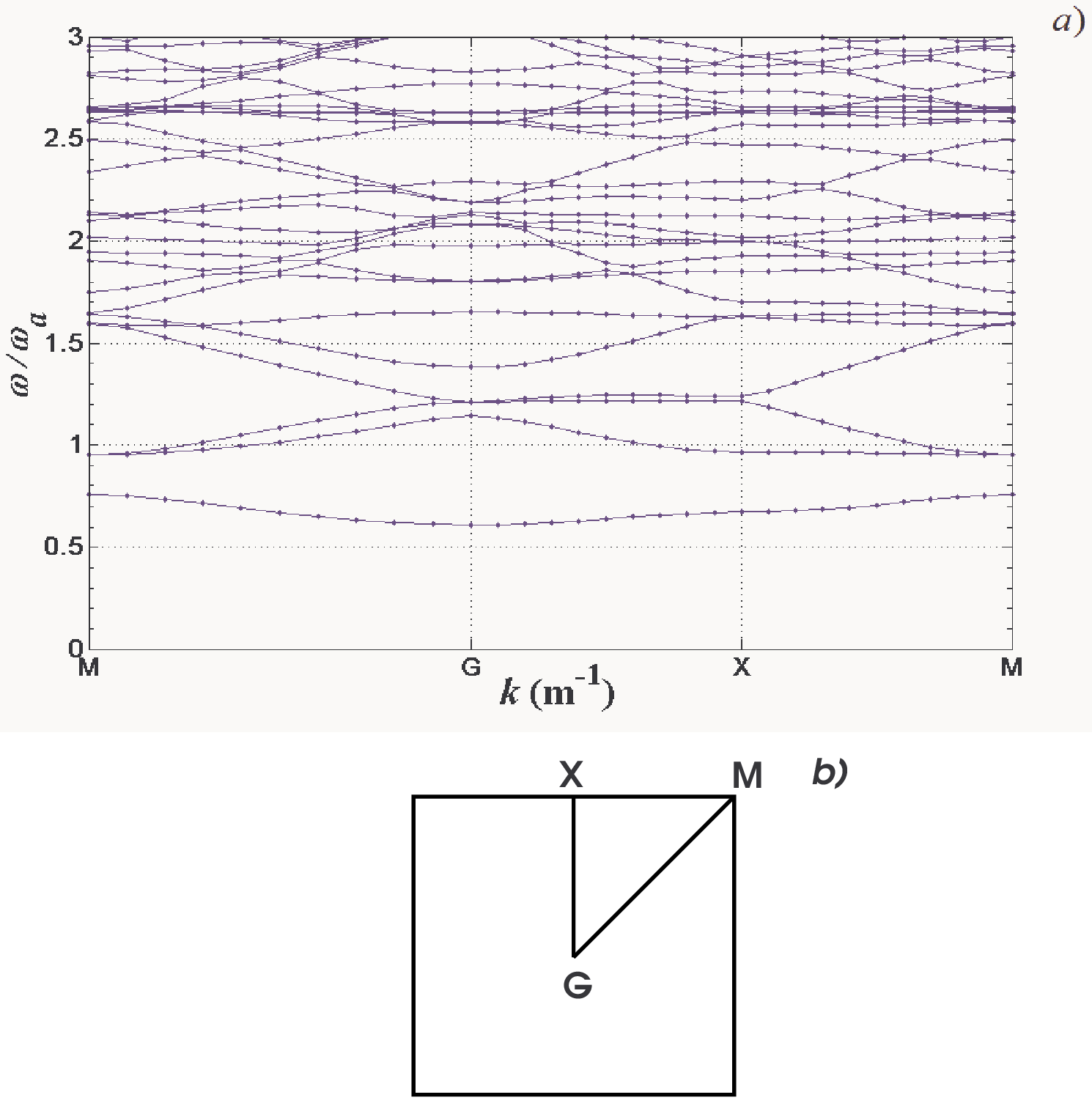}
\caption{a) Band structure of graphene based 2D square photonic
crystal of cylinder array arranged in a square lattice. The
cylinders consist of metamaterial stack of graphene monolayer discs
separated by the dielectric discs. The filling factor $f = 0.3927$.
$M$, $G$, $X$, $M$ are points of symmetry in the first (square)
Brillouin zone. b) The first Brillouin zone of the 2D photonic
crystal. } \label{band}
\end{figure}

\begin{figure}
\includegraphics[angle=90,width = 3.8in]{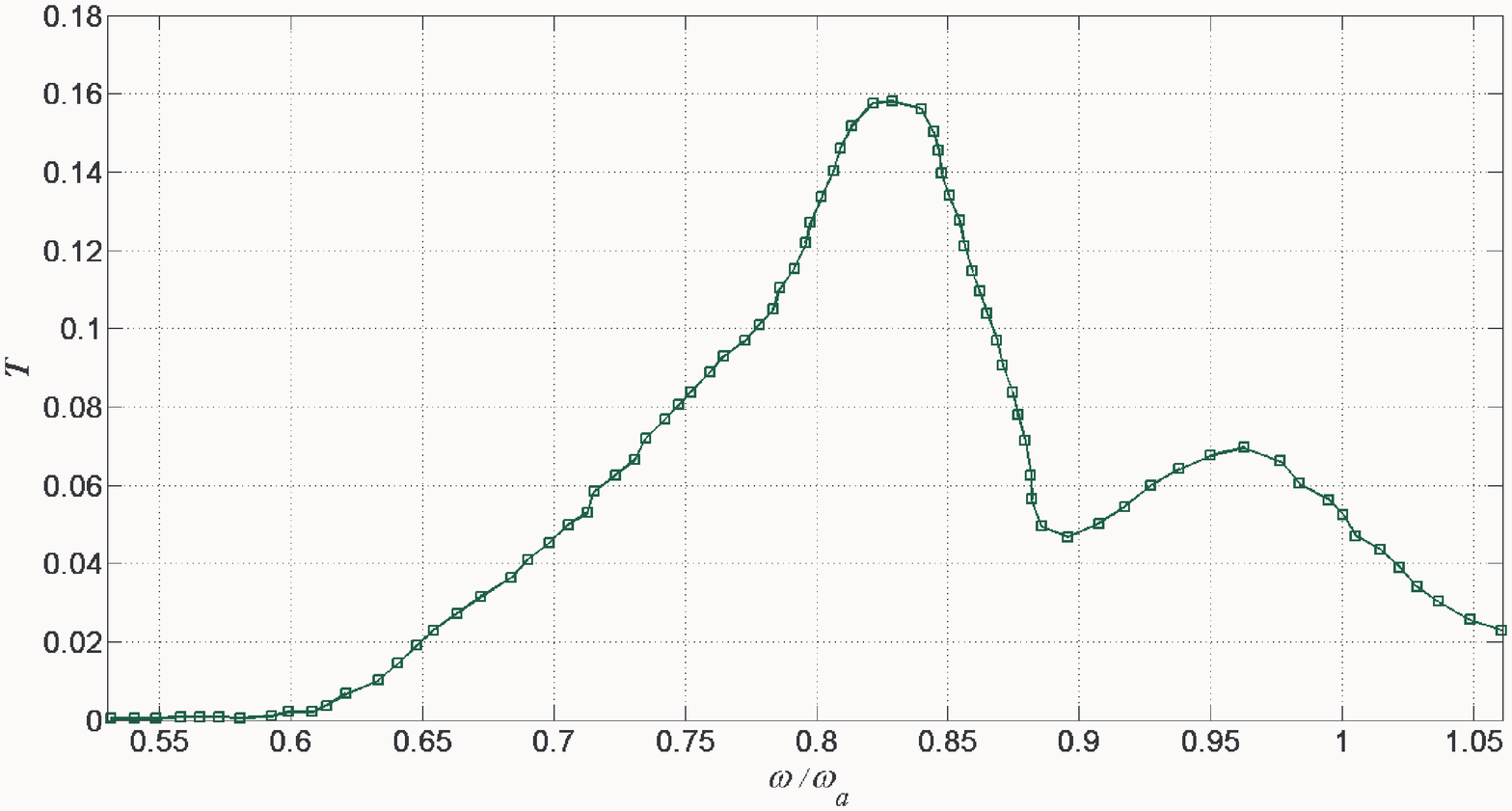}
\caption{The transmittance $T$ spectrum of graphene based 2D
photonic crystal.   }\label{trans}
\end{figure}

To illustrate the effect let us, for example, consider the 2D square
lattice formed by the graphene based metamaterial embedded in the
dielectric. The photonic band structure for the graphene based 2D
photonic crystal with the array of cylinders arranged in a square
lattice with the filling factor $f = 0.3927$ is presented in
Fig.~\ref{band}. The cylinders consist of the metamaterial stacks of
alternating graphene and dielectric discs. The period of photonic
crystal is $a = 25 \ \mathrm{\mu m}$, the diameter of discs is $D =
12.5 \ \mathrm{\mu m}$, the width of the dielectric layers $d =
10^{-3} \ \mathrm{\mu m}$ . Thus the lattice frequency is
$\omega_{a} = 2 \pi c/a = 7.54 \times 10^{13} \ \mathrm{rad/s}$. The
results of the plane wave calculation for the graphene based
photonic crystal are shown in Fig.~\ref{band}, and the transmittance
spectrum obtained using the Finite-Difference Time-Domain (FDTD)
method~\cite{Taflove}
 is presented in Fig.~\ref{trans}. Let us mention that plane wave
computation has been made for extended photonic crystal, and FDTD
calculation of the transmittance have been performed for five
graphene layers.  A band gap is clearly apparent in the frequency
range $0 < \omega < 0.6$ and $0.75 < \omega < 0.95$ in units of
$2\pi c/a$. The first gap is originated from the electronic
structure of the doped graphene, which prevents absorbtion at
$\hbar\omega < 2\mu$ (see also Eq.~(\ref{cond})). The photonic
crystal structure manifests itself in the dependence of the lower
photonic band on the wave vector $k$. In contrast, the second gap
$0.75 < \omega < 0.95$ is caused by the photonic crystal structure
and dielectric contrast. 

According to Fig.~\ref{trans}, the transmittance $T$ is almost zero
for the frequency lower than $0.6 \omega_{a}$, which corresponds to
the first band gap shown in Fig.~\ref{band}.   The second gap in
Fig.~\ref{band} (at the point $G$) corresponds to $\omega = 0.89
\omega_{a}$, and it also corresponds to the transmittance spectrum
minimum on Fig.~\ref{trans}.

Let us mention that at $\hbar \omega < 2\mu$ the
dissipation of the electromagnetic wave in graphene is suppressed.
In the long wavelength (low frequency) limit, the skin penetration
depth is given by $\delta_{0}(\omega) = c/Re \left[2\pi \omega
\sigma_{g}(\omega)\right]^{1/2}$~\cite{Landau}. According to
Eq.~(\ref{cond}), $Re [\sigma_{g}(\omega < 2\mu)] = 0$, therefore,
$\delta_{0}(\omega) \rightarrow + \infty$, and the electromagnetic
wave penetrates along the graphene layer without damping. For the
carrier densities $n_{0} = 10^{11}
 \  \mathrm{cm^{-2}}$ the chemical potential is  $\mu = 0.022 \ \mathrm{eV}$~\cite{Falkovsky_prb}, and
 for the frequencies $\nu < \nu_{0} = 10.42 \  \mathrm{THz}$ we have $Re [\sigma_{g}(\omega)] =
 0$  at $\omega \gg 1/\tau$  the electromagnetic wave
penetrates along the graphene layer almost without damping,
 which makes the graphene multilayer based photonic crystal to be
 distinguished from the metallic  photonic crystal, where the electromagnetic
 wave is essentially damped.
As a result, the graphene-based photonic crystals can have the sizes much larger than the  metallic photonic crystals.
 The scattering of the electrons on the impurities can
result in non-zero $Re [\sigma_{g}(\omega)]$, which can cause the
dissipation of the electromagnetic wave. Since the electron mobility
in graphene can be much higher than in typical semiconductors, one
can expect that the scattering of the electrons on the impurities
does not change the results significantly.

The physical properties of graphene-based photonic crystals are
different from the physical properties of other types of photonic
crystals, since the dielectric constant of graphene has the unique
frequency dependence~\cite{Falkovsky_prb,Falkovsky_conf}.
 According to the results presented above, the
graphene-based photonic crystal has completely different photonic
band structure in comparison  to the photonic crystals based on the other
materials. The photonic band structure of the photonic
crystal with graphene multilayer  can be tuned by changing the
distance $d$ between graphene discs  in the r.h.s. of
Eq.~(\ref{dielmult}). The photonic band structure of the
graphene-based photonic crystals can also be controlled by the
doping, which determines the chemical potential $\mu$ entering the
expressions for the conductivity and dielectric constant of graphene
multilayer~(\ref{cond}).

Comparing the photonic band structure for graphene-based photonic
crystal presented in Fig.~\ref{band} with the
dielectric~\cite{Joannopoulos2}, metallic~\cite{Maradudin,Kuzmiak},
semiconductor~\cite{Maradudin} and
superconductor-based~\cite{Berman1,Lozovik} photonic crystals, we
conclude that only graphene- and superconductor-based photonic
crystals have essential photonic band gap at low frequencies
starting $\omega = 0$, and the manifestation of the gap in the
transmittance spectra is almost not suppressed by the damping
effects. Therefore, only graphene-based and superconducting photonic
crystals can be used effectively as the frequency filters and
waveguides in low-frequency for the far infrared region of spectrum,
while the devices based on the dielectric photonic crystals can be
used only in the optical region of electromagnetic waves spectrum.
The graphene based-photonic crystal can be used at room
temperatures, while the superconductor-based photonic crystal can be
used only at low temperatures below the critical temperature
$T_{c}$, which is about $90 \  \mathrm{K}$ for the $\mathrm{YBCO}$
superconductors.

\emph{In summary}, the graphene-based photonic crystal presented in
this Letter is the novel type of photonic crystal. The
frequency band structure of a
 2D photonic crystal with the square lattice of the metamaterial stacks of the alternating graphene and dielectric discs is obtained.
The electromagnetic wave transmittance of such photonic crystal is
calculated.
The graphene-based photonic crystals have the following advantages that distinguish them from the other types of photonic crystals.
 They can be used as the frequency filters and waveguides for the far-infrared region of spectrum at the wide range of the temperatures
 including the room temperatures. The photonic band structure of the graphene-based photonic crystals can be controlled by  changing the thickness of the dielectric
 layers between the graphene discs and by the doping.
 The sizes of the graphene-based photonic crystals can be much larger than the sizes of metallic photonic crystals due   to the small dissipation of the electromagnetic wave.
 Let us also mention that above for simplicity we assume  that the dielectric material is
the same between the graphene disks and between the stacks. This
assumption has some technological advantage for the most easier
possible experimental realization of the graphene-based photonic
crystal.

\acknowledgments

O.~L.~B., V.~S.~B., R.~Ya.~K. were supported by PSC CUNY grant
63443-00 41 and grant 61383-00 39; Yu.~E.~L. and A.~A.~K. were
supported by RFBR and RAS programs.


\end{document}